\documentstyle[12pt]{article}

\begin{document}
\begin{center}
{\Large \bf Michelson-Morley experiment within the
quantum mechanics framework}

\bigskip

{\large D.L.~Khokhlov}
\smallskip

{\it Sumy State University, Ukraine\\
E-mail: dlkhokhl@rambler.ru}
\end{center}

\begin{abstract}

It is revisited the Michelson-Morley experiment within the
quantum mechanics framework. One can define the wave function of
photon in the whole space at a given moment of time. The phase
difference between the source and receiver is a distance between
the source and receiver at the time of reception hence it does not
depend on the velocity of the frame. Then one can explain the null
result of the Michelson-Morley experiment within the quantum
mechanics framework.

\end{abstract}

The Maxwell-Lorentz equations describe electromagnetic field as a
wave propagating with the velocity $c$. It is reasonable to think
that electromagnetic wave propagates with the velocity $c$ with
respect to a privilege frame. If some frame moves with the
velocity $v$ with respect to a privilege frame then one can
expect that electromagnetic wave propagates with the velocity
$\vec{c}-\vec{v}$ with respect to the moving frame.
The Michelson-Morley experiment was suggested to determine
the velocity of electromagnetic wave with respect to a moving
frame with the earth being taken as a moving frame.
However the Michelson-Morley experiment~\cite{Pau} yielded the
null result. The special relativity~\cite{Pau} explains
the null result of the Michelson-Morley experiment
with the Lorentz transformation for coordinates of space and time.
Electromagnetic field is a quantum object.
Below we shall revisit the Michelson-Morley experiment
within the quantum mechanics framework.

Consider electromagnetic field within the Newtonian framework.
Consider electromagnetic field as a wave with the vector potential
$\vec{A}$ in the Euclidean space and time of a privilege frame.
We shall take the cosmic microwave background radiation
(CMB)~\cite{Lin} as a privilege frame.
The Maxwell-Lorentz equations for the electromagnetic wave are
given by~\cite{Lan}
\begin{equation}
\triangle\vec{A}-\frac{1}{c^2}
\frac{\partial^2\vec{A}}{\partial t^2}=0
\label{eq:A}
\end{equation}
where $c$ is the velocity of light.
One can represent the solution of eq.~(\ref{eq:A}) as a plane
monochromatic wave
\begin{equation}
\vec{A}=\vec{A_0}e^{-i\phi}
\label{eq:A2}
\end{equation}
with the phase
\begin{equation}
\phi=\omega t-kr
\label{eq:phi}
\end{equation}
where $\omega$ is the frequency, $k$ is the wave vector.

In the quantum mechanics~\cite{Dir}, one can consider
electromagnetic field as a bunch of photons with the momentum and
energy given by respectively
\begin{equation}
p=\hbar k \qquad {\mathcal{E}}=\hbar\omega
\label{eq:pe}
\end{equation}
where $\hbar$ is the Planck constant.
One can conceive the photon as a particle exhibiting
wave behaviour. In the quantum mechanics the wave given by
eq.~(\ref{eq:A2}) is thought of as a wave function of photon
associated with a single photon.
For photons the Heisenberg uncertainty principle holds true
\begin{equation}
\Delta p \Delta r\geq\frac{\hbar}{2}\qquad
\Delta{\mathcal{E}}\Delta t\geq\frac{\hbar}{2}.
\label{eq:Hei}
\end{equation}
In view of eq.~(\ref{eq:pe}), the Heisenberg uncertainty principle
restricts the wave function of photon in the space and time.

Consider the photon as a particle with the momentum $p$ at the
time $t$. Introduce the wave function of photon with
the wave vector $k=p/\hbar$.
In the stationary state the momentum of photon is fixed.
Then the uncertainty in momentum is $\Delta p=0$,
the uncertainty in the wave vector is $\Delta k=\Delta p/\hbar=0$.
From eq.~(\ref{eq:Hei}) it follows the uncertainty in space
coordinate $\Delta r=\infty$.
This means one cannot specify the space coordinate hence one can
consider the wave function of photon with the wave
vector $k=p/\hbar$ in the whole space at the time $t$.
One can describe the wave function of photon by
eq.~(\ref{eq:A2}) like the classical wave.
However one cannot treat electromagnetic field as a classical wave
and conceive it as a fluid consisting of a number of photons.
One can treat electromagnetic field as a wave function
accompanying to a single photon.

Consider propagation of a photon between the source and receiver.
We shall regard the case when the region of propagation
of photon is much more than the wave length of photon
$\Delta r\gg\lambda=1/k$. Then one can think of the photon as a
point-like quasi-classical particle propagating with the velocity
$c$ with respect to the CMB frame.
In view of the above reasoning one can define the wave function of
photon in the whole space at the time of reception.
Then the phase difference between the source and
receiver is a distance between the source and receiver at the time
of reception
\begin{equation}
\Delta\phi=\phi_r(t_r)-\phi_s(t_r)=k[r_r(t_r)-r_s(t_r)].
\label{eq:dphi}
\end{equation}
We come to the definition of phase within the quantum mechanics
framework. In the classical physics the phase is defined through
the space coordinate of the source at the time of emission
$r_s(t_e)$. In the quantum mechanics the phase is defined through
the space coordinate of the source at the time of reception
$r_s(t_r)$. The quantum mechanics definition of phase proceeds
from the fact that the wave function of photon is associated with
a single photon hence is defined in the whole space at the time of
reception.

Consider the Michelson-Morley experiment in a frame moving with
the velocity $v$ with respect to the CMB frame.
Suppose that electromagnetic field (photon) moves with the
velocity $c$
with respect to the CMB frame independently of the source
(receiver). Then the travel time is a function of the velocity of
the frame with the maximum difference of travel time between two
legs for two-way travel being $\Delta t=(l/c)(v^2/c^2)$
where $l$ is the length of the leg.
According to the quantum mechanics~\cite{Dir} a single photon
interferes with itself. The wave function of photon is a
superposition of the waves specified along two different legs
with the photon as a particle moving along one of the legs.
In view of eq.~(\ref{eq:dphi}), the phase difference between the
source and receiver is a distance between the source and receiver
at the time of reception. If the lengths of the legs are not the
same there is a phase shift between two waves specified along two
different legs $\Delta\phi=k(l_2-l_1)$.
When determining the distance between the source and
receiver at one and the same moment of time it does not depend on
the velocity of the frame.
Hence the phase difference between the source and receiver does
not depend on the velocity of the frame.
Hence there is no phase shift due to the velocity of the frame
between two waves specified along two different legs.
Thus one can explain the null result of the Michelson-Morley
experiment within the quantum mechanics framework without invoking
the Lorentz transformation.

According to the special relativity~\cite{Pau} coordinates of
space and time follow the Lorentz transformation (LT),
$r^\prime=r{\mathrm{LT}}$, $t^\prime=t{\mathrm{LT}}$
while the wave vector and frequency of electromagnetic wave follow
the inverse Lorentz transformation $k^\prime=k{\mathrm{LT}}^{-1}$,
$\omega^\prime=\omega{\mathrm{LT}}^{-1}$.
Here the non-primed values are the proper ones while the primed
values are the apparent ones. The phase
of electromagnetic wave is Lorentz invariant
$\phi^\prime=\omega^\prime t^\prime-k^\prime r^\prime=
\omega t-kr=\phi$ that explains
the null result of the Michelson-Morley experiment.

Explanation of the null result of the Michelson-Morley experiment
within the quantum mechanics framework allows Galilean invariance
of the electromagnetic wave. One may treat the phase given by
eq.~(\ref{eq:dphi}) as Galilean invariant. That
is both observers in a privilege and in a moving frames determine
the same space coordinate difference and the same wave vector.
The wave vector of the electromagnetic wave emitted in a moving
frame is Lorentz shifted in a privilege frame (Doppler effect).
Assume that electromagnetic wave behaves in a different way
under propagation of the electromagnetic wave and under
interaction with the source. Under propagation electromagnetic
wave is Galilean invariant while under interaction with the source
is Lorentz invariant.
Then we may apply the conventional special relativity
under interaction of the electromagnetic wave with the source.
After emission
the wave vector of the electromagnetic wave is the same for both
observers in a privilege and in a moving frames. That is
Galilean invariance of the phase
under propagation of the electromagnetic wave means
$\phi=k^\prime r={\mathrm{Gal\ inv}}$.

So putting the quantum mechanics definition of the wave function
of photon one can explain the null result of the Michelson-Morley
experiment without invoking the Lorentz transformation.
It is reasonable to assume that
under propagation electromagnetic wave is Galilean invariant.
Explanation of the Doppler effect needs the Lorentz
transformation. It is reasonable to assume that under interaction
with the source electromagnetic wave is Lorentz invariant.

\end{document}